\documentclass{article}
\usepackage{latexsym}
\usepackage{amssymb,amsmath}
\usepackage{amsthm}
\usepackage{graphicx}
\usepackage{fullpage}
\usepackage{url}

\def\leq{\leqslant}
\def\geq{\geqslant}

\newcommand{\R}{\mathbb{R}}

\newcommand{\s}{\mathbb{S}}

\newcommand{\B}{\mathcal{S}}

\newcommand{\Dirs}{\mathcal{D}}
\newcommand{\Lines}{\mathcal{L}}
\newcommand{\DS}{\mathcal{D}_{\F}}
\newcommand{\K}{\mathcal{K}}
\newcommand{\Ko}{\K^{\circ}}
\newcommand{\F}{\mathcal{F}}

\newcommand{\h}{\mathcal{H}}
\makeatletter
\def\INT{\mathop{\operator@font int}\nolimits}
\makeatother

\newtheorem {theo}         {Theorem}

\newtheorem {lemma} [theo] {Lemma}
\newtheorem {prop}  [theo] {Proposition}
\newtheorem {cor}   [theo] {Corollary}
\newtheorem {pb}           {Problem}

\title{Helly-Type Theorems for Line Transversals to Disjoint Unit Balls\thanks{Andreas Holmsen was supported by the Research Council of Norway, prosjektnummer 166618/V30. Otfried Cheong and Xavier Goaoc acknowledge support from the French-Korean Science and Technology Amicable Relationships program (STAR).}}
\date{}
\author{Otfried Cheong%
  \thanks{Division of Computer Science, KAIST, Daejeon, South Korea.
    Email:~\textrm{otfried@kaist.ac.kr}.}
  \and
  Xavier Goaoc%
  \thanks{LORIA - INRIA Lorraine, Nancy, France.
    Email:~\textrm{goaoc@loria.fr}.}
  \and
  Andreas Holmsen%
  \thanks{Department of Mathematics, University of Bergen, Bergen,
    Norway. Email:~\textrm{andreash@mi.uib.no}.}
  \and
  Sylvain Petitjean%
  \thanks{LORIA - CNRS, Nancy, France.
    Email:~\textrm{petitjea@loria.fr}.}
}

\begin{document}

\maketitle

\begin{abstract}
  We prove Helly-type theorems for line transversals to disjoint unit
  balls in~$\R^{d}$. In particular, we show that a family of $n \geq
  2d$ disjoint unit balls in $\R^d$ has a line transversal if, for
  some ordering $\prec$ of the balls, any subfamily of $2d$ balls
  admits a line transversal consistent with $\prec$. We also prove
  that a family of $n \geq 4d-1$ disjoint unit balls in $\R^d$ admits
  a line transversal if any subfamily of size $4d-1$ admits a
  transversal.
\end{abstract}

\section{Introduction}

Helly's celebrated theorem, published in 1923, states that a finite family of
convex sets in $\R^d$ has non-empty intersection if and only if any subfamily of
size at most $d+1$ has non-empty intersection. Subsequent results of similar
flavor (that is, if every subset of size $k$ of a set $\B$ has property $\cal P$
then $\B$ has property $\cal P$) have been called \emph{Helly-type theorems} and
the minimal such $k$ is known as the associated \emph{Helly number}. Helly-type
theorems and tight bounds on Helly numbers have been the object of active research
in combinatorial geometry. In this paper, we investigate Helly-type theorems for
the existence of line transversals to a family of objects, i.e. lines that
intersect every member of the family.

\paragraph{History.}
The earliest Helly-type theorems in geometric transversal theory
appeared about five decades ago. In 1957, Hadwiger~\cite{h-uegt-57}
showed that an ordered family $\B$ of compact convex sets in the
plane admits a line transversal if every triple admits a line
transversal compatible with the ordering. (Note that a line transversal
to $\B$ may not respect the ordering on $\B$; to prove the existence
of a line transversal that respects the ordering on~$\B$ one needs the
assumption that any \emph{four}-tuple admits an order-respecting line
transversal.) In what follows, we shall talk about a Hadwiger-type theorem when
the family of objects under consideration is ordered.

The same year, Danzer~\cite{Danzer57} proved the following result
concerning families of pairwise disjoint unit discs in the plane: if
such a family consists of at least 5 discs, and if any 5 of these
discs are met by some line, then there exists a line meeting all the
discs of the family. This answered a question of
Hadwiger~\cite{Hadwiger55}, who gave an example (5 circles, almost
touching and with centers forming a regular pentagon) which shows
that~5 cannot be replaced by~4. Gr\"unbaum~\cite{grunbaum58} showed
that the same result holds if ``unit disc'' is replaced by ``unit
square'', and conjectured that the result holds for families of
disjoint translates of any compact convex set in the plane. This
long-standing conjecture was finally proved by
Tverberg~\cite{tverberg89}. A weaker form of the conjecture which
assumed 128 instead of 5 had been established earlier by
Katchalski~\cite{katch86}.

Danzer~\cite{Danzer57} conjectured that Helly-type theorems exist for
line transversals to disjoint unit balls in arbitrary dimension. The
first positive result was obtained by Hadwiger~\cite{h-wo-57} for the
case of families of ``thinly distributed'' balls, where the distance
between any two balls is at least the sum of their radii. This result
was extended by Ambrus et al.~\cite{abf-httdub-06} to disjoint unit
balls, in arbitrary dimension, the centers of which are at distance at
least $2\sqrt{2+\sqrt{2}}$. Danzer's conjecture for three-dimensional
disjoint unit balls, without additional assumption on their
distribution, was only settled in 2001 by Holmsen et
al.~\cite{hkl-httlt-03}. It should be stressed that in dimension three
(and higher), neither Hadwiger nor Helly-type theorems exist for line
transversals to general convex objects, not even for translates of a
convex compact set~\cite{hm-nhtst-04}.

In his paper~\cite{Danzer57}, Danzer also asked whether the Helly
number for line transversals to disjoint unit balls in $\R^d$ is a
strictly increasing function of $d$. The only known lower bound is the
planar example of Hadwiger~\cite{Hadwiger55}. This number was proved
to be at most $d^2$ for thinly distributed balls in $\R^d$ by
Hadwiger~\cite{h-wo-57}, a bound improved to~$2d-1$ by
Gr\"unbaum~\cite{g-ctfs-60} using the topological Helly theorem. For
disjoint unit balls in dimension three, Holmsen et
al.~\cite{hkl-httlt-03} proved bounds of respectively~$12$ and~$46$
for the Hadwiger-type and Helly-type theorems, which were later improved
to~$12$ and~$18$ by Cheong et al.~\cite{cgn-gpdus-05}.

 
\medskip
We refer the reader to the recent survey by Wenger~\cite{w-httgt-04}
for a broader discussion of geometric transversal theory.

\paragraph{Our results.}
In this paper we complete the proof of Danzer's conjecture. More
precisely, we show that Helly-type theorems exist for line
transversals to families of \emph{pairwise-inflatable} balls in
$\R^{d}$.  A family $\F$ of balls in $\R^{d}$ is called
pairwise-inflatable if for every pair of balls $B_{1}, B_{2}
\in \F$ we have $\gamma^{2} > 2(r_{1}^{2} + r_{2}^{2})$, where $r_{i}$
is the radius of $B_{i}$, and $\gamma$ is the distance between their
centers. A family of disjoint unit balls is pairwise-inflatable, since
$\gamma^{2} > 2(r_{1}^{2} + r_{2}^{2})$ implies $\gamma >
r_{1}+r_{2}$ when $r_1=r_2$, and so is a family of balls that is
``thinly distributed'' in Hadwiger's sense.  Pairwise-inflatable
families of balls are not only more general than families of disjoint
congruent balls but allow to generalize most of our proofs obtained in
three or four dimensions to arbitrary dimension; the key property,
which we prove in this paper, is that the set of pairwise-inflatable
families is closed under intersection with affine subspaces, unlike
the set of families of disjoint congruent balls.


An order-respecting line transversal to a subset of an ordered family
is a line transversal that respects the order induced by the family on
that subset. An ordered family $\F$ of pairwise-inflatable balls is said to have
property $(OR)T$ if it admits a (order-respecting) line
transversal. If every $k$ or fewer members of $\F$ admit a (order-respecting) line
transversal then $\F$ is said to have property 
$(OR)T(k)$. Our first main result requires that the line transversals
to the subfamilies induce consistent orderings:

%
\begin{theo}\label{Hadwiger}
For any ordered family of pairwise-inflatable balls in
  $\R^d$, $ORT(2d)$ implies $T$ and $ORT(2d+1)$ implies $ORT$.
\end{theo}
We then remove the condition on the ordering at the cost of increasing
the Helly number to~$4d-1$ and restricting ourselves to disjoint
unit balls:
\begin{theo}\label{Helly-type}
 For any family of disjoint unit balls in $\R^d$, $T(4d-1)$ implies $T$.
\end{theo}

%
%
%
%
Our results are thus both qualitative and quantitative: we generalize
Danzer's result to arbitrary dimension and prove that the Helly number
grows at most linearly with the dimension. We build on the work of
Holmsen et al.~\cite{hkl-httlt-03} who obtained results similar to
Theorems~\ref{Hadwiger} and~\ref{Helly-type} for disjoint unit balls
in three dimensions, albeit with larger bounds on Helly numbers ($12$ and $46$
instead of $6$ and $11$, respectively). A previous version of this
paper, also restricted to disjoint unit balls in three dimensions,
appeared in the Symposium on Computational
Geometry~2005~\cite{cgh-hhtdus-05}.

\paragraph{Paper outline.}
To prove Theorem~\ref{Hadwiger}, we start with a family of balls
having property $ORT(2d)$ and continuously shrink them until that
property no longer holds, following Hadwiger's
approach~\cite{h-uegt-57}. Before the set of order-respecting line
transversals to a $2d$-tuple of balls disappears, it first reduces to
a single line (Corollary~\ref{empty-interior-anyD}) and this line is
an isolated line transversal to $2d-1$ of the balls
(Proposition~\ref{pinning}). That line has then to be a line
transversal to the whole family and Theorem~\ref{Hadwiger} follows;
considerations on geometric permutations yield
Theorem~\ref{Helly-type}.

Proving the two properties mentioned above
(Corollary~\ref{empty-interior-anyD} and Proposition~\ref{pinning}) is
elementary in the plane but requires considerably more work in higher
dimension. Our proofs rely on Proposition~\ref{convex-inflatable}, the
cornerstone of this paper, which shows that the directions of
order-respecting line transversals to a family of pairwise-inflatable
balls form a strictly convex subset of $\s^{d-1}$. This directly
implies Corollary~\ref{empty-interior-anyD} and yields that
order-respecting line transversals form a contractible set in line
space. From there, a well-known topological analogue of Helly's
theorem (Theorem~\ref{thm:topological-helly}) leads to a weaker
version of Theorem~\ref{Hadwiger} sufficient to prove
Proposition~\ref{pinning}.

%
%
\section{Preliminaries}\label{sec:prel}

\paragraph{Transversals.}
Let $\F$ be a finite family of disjoint compact convex sets $\F$ in
$\R^{d}$ with a given linear order~$\prec_{\F}$.  We will call $\F$ a
\emph{sequence} to stress the existence of this order.  A \emph{line
transversal} to a \emph{sequence} $\F$ is an oriented line that
intersects all the objects of~$\F$ \emph{in the order prescribed
by~$\prec_{\F}$}.  A line transversal is \emph{strict} if it
intersects the \emph{interior} of each object in~$\F$.

For a sequence $\F$, let $\K(\F)\subset \s^{d-1}$ denote the set of
directions of line transversals to $\F$.  That is, a direction vector
$v \in \s^{d-1}$ is in $\K(\F)$ if there is a line transversal to $\F$
with direction~$v$.  Note that the direction vector of a line
transversal determines the order in which it intersects a family of
disjoint convex objects.  Thus, if sequences $\F_1$ and $\F_2$ are two distinct
orderings of the same collection of objects, then $\K(\F_1)$ and
$\K(\F_2)$ are disjoint.  We will call $\K(\F)$ the 
\emph{cone of directions} of $\F$.  Similarly, let $\Ko(\F)$ be the
set of directions of \emph{strict} line transversals to~$\F$.

Note that all our line transversals must respect a given order.  Only
in Section~\ref{hadwiger-and-helly} will we consider line transversals
without order restriction.  For clarity, let us call such a line
transversal an \emph{unordered} line transversal.

We consider the natural topology over the set of oriented lines in
$\R^d$: $U$ is a neighborhood of a line $\ell$ if and only if for some
$\delta > 0$ it contains all lines $\ell'$ such that the shortest
distance between $\ell$ and $\ell'$ and the angle between their
direction vectors are both less than~$\delta$. An \emph{isolated} line
transversal to a family of objects $\F$ is an isolated point of the
set of line transversals to $\F$, that is, a line transversal $\ell$
which is a connected component of the line transversals to $\F$. 

Given a ball $A$ and a direction $v$ in $\R^d$, we denote by $P_v(A)$
the $(d-1)$-dimensional ball obtained by projecting $A$ orthogonally
on an hyperplane with normal~$v$. Observe that a sequence of balls
$\F$ has a line transversal with direction $v$ if and only if the
balls $P_v(\F) := \{P_v(A)\mid A\in \F\}$ have non-empty
intersection. Similarly, $\F$ has a strict line transversal with
direction $v$ if and only if the intersection of $P_v(\F)$ has
non-empty interior.

\paragraph{Inflatable balls.}
A collection $\F$ of balls in $\R^d$ is called
\emph{pairwise-inflatable} if for every two balls $B_{1}, B_{2} \in
\F$ we have $\gamma^{2} > 2(r_{1}^{2} + r_{2}^{2})$, where $r_{i}$ is the
radius of $B_{i}$, and $\gamma$ is the distance between their centers.
Note that for balls of equal radius, this condition only enforces that
they are disjoint (and so any family of disjoint congruent balls is
pairwise-inflatable).  The more unequal the radius of the balls,
however, the stronger the distance constraint. At the limit, when $r_{1}
= 0$, the constraint is $\gamma > \sqrt{2} r_{2}$.  Pairwise-inflatability
is less restrictive than Hadwiger's notion of ``thinly distributed'' balls, which
can be defined as $\gamma^{2} > 4(r_{1} + r_{2})^{2}$ for each pair of balls.

The class of families of pairwise-inflatable balls is closed under
intersection with affine subspaces (as proved in Lemma~\ref{lem:pi-closed}).  This
property (which does not hold for unit-radius balls) will allow us to
carry results proved in three dimensions over to~$\R^{d}$.

\paragraph{Topological machinery.}
We use a few notions from topology that we now review (these can be
found, for instance, in the introductory chapter of Matou{\v s}ek's
book~\cite{m-ubut-03}). Given a topological space $A$ and a subset $B
\subset A$, $B$ is a \emph{deformation retract} of $A$ if there exists
a continuous map $F: A\times[0,1] \rightarrow A$ such that
\[
\left\{
\begin{array}{l}
  F(a,0) = a \hbox{ for any } a\in A\\
  F(b,t) = b \hbox{ for any $b\in B$ and $t \in [0,1]$} \\
  F(a,1) \in B \hbox{ for any } a\in A
\end{array}
\right.
\]
Two topological spaces $A, B$ are \emph{homotopy equivalent} if there
exists a third space $C$ such that both $A$ and $B$ are deformation
retracts of~$C$.  A space that is homotopy equivalent to a single
point is said to be \emph{contractible}. A homology cell is a
non-empty set with trivial homology, e.g. a point. Since homology is
invariant under homotopy equivalence, any contractible space is a
homology cell. A generalization of Helly's theorem based on topology
instead of convexity was originally given by Helly
himself~\cite{h-usvam-30}.  We will use a version proved by Debrunner
using modern tools (singular homology)~\cite{d-httdb-70}, as it allows
us to work with open sets.

\begin{theo}[Topological Helly Theorem~\cite{d-httdb-70}]
  \label{thm:topological-helly}
  Let $\{X_{j}\}_{j \in J}$ be a finite family of open subsets of
  Euclidean $d$-space $\R^d$ such that the intersection $X_{j_{1}} \cap
  \dots \cap
  X_{j_{r}}$ of each $r$ sets of this family is nonempty for $r \leq
  d+1$ and is even a homology cell for $r \leq d$. Then
  $\bigcap_{j\in J}X_{j}$ is a homology cell.
\end{theo}

\noindent In fact, we only use a weaker version of this theorem where ``homology cell'' is replaced by ``contractible''.

\paragraph{Compatible directions.}
Let $\Dirs$ be a set of directions in $\R^d$ completely contained in
the interior of a hemisphere of $\s^{d-1}$, and let $\Lines(\Dirs)$ be
the set of lines with directions in~$\Dirs$. We parametrize
$\Lines(\Dirs)$ as a subset of $\R^{2d-2}$, using the points of
intersection of a line $\ell \in \Lines(\Dirs)$ with two parallel
hyperplanes that are not parallel to any direction in~$\Dirs$.  Our
aim is to apply the Topological Helly Theorem to sets of line
transversals to pairwise-inflatable balls. Unfortunately, such sets
are not necessarily homology cells, and may in fact even be
disconnected: two lines intersecting disjoint objects in different
orders cannot be in the same connected component of transversals to
these objects. We overcome this difficulty by restricting the set of
directions that we allow for transversals.  For a sequence $\F$ of
pairwise-inflatable balls in $\R^{d-1}$, let
\[
U(\F) := \bigl\{ c(Y) - c(X) \bigm| X,Y \in \F; \; X \prec_{\F} Y
\bigr\},
\]
where $c(X)$ denotes the center of ball~$X$. Let $\DS$ be the set of
directions making a positive dot-product with each $u \in U(\F)$.
Note that $\DS$ is an open convex set on the sphere of directions
$\s^{d-1}$. Clearly a line transversal $\ell \in \Lines(\DS)$ for a
subset $\F' \subset \F$ respects the order on~$\F'$. Such a line
transversal is called a transversal to $\F'$ \emph{compatible} with
$\F$.

\section{The cone of directions is strictly convex}

We now establish the cornerstone of this paper, a generalization of
the first lemma by Holmsen et al.~\cite{hkl-httlt-03} to arbitrary
dimension:

\begin{prop}\label{convex-inflatable}
  Let $\F$ be a sequence of pairwise-inflatable balls in $\R^d$. Then
  $\K(\F)$ is strictly convex.
\end{prop}

The proof of this proposition is based on Lemma~\ref{QAB-4D}, which
shows that some well-chosen fibers over $1$-dimensional slices of the
cone of directions of unit balls in $\R^4$ are convex.  We also need
some properties of families of pairwise-inflatable balls. We start by
showing that this class is closed under intersection with affine
subspaces.

\begin{lemma}
  \label{lem:pi-closed}
  Let $\F$ be a family of pairwise-inflatable balls in $\R^{d}$, and
  let $E$ be an affine subspace of dimension $k < d$.  Then $\F' := \{
  B \cap E \mid B \in \F \}$ is a family of pairwise-inflatable balls
  in~$E$.
\end{lemma}
\begin{proof}
We prove the claim for $k = d - 1$ and the lemma follows by induction. Let $B_{1},
B_{2} \in \F$ with respective radii $r_1$ and $r_2$ and centers at distance
$\gamma$ apart. Since $\F$ is pairwise-inflatable we have $\gamma^2 > 2
(r_1^2+r_2^2)$. For $i = 1, 2$ let $B'_{i} = B_{i} \cap E$, $\rho_i$ denote the
radius of $B'_i$ and $\delta_i$ be the distance between the center of $B_i$ and
that of $B'_i$. First, observe that  
\[
\gamma^{2} \leq \Delta^{2}+(\delta_{1} + \delta_{2})^{2},
\]
where $\Delta$ is the distance between the centers of $B_1'$ and
$B_2'$.  If $E$ separates the centers of $B_1$ and $B_2$ the equality
holds. If $E$ does not separate the
centers, then replacing $B_2$ by its mirror image with respect to~$E$
increases $\gamma$ while leaving all other quantities unchanged, hence
the inequality. Then from $(\delta_{1}-
\delta_{2})^{2} \geq 0$ we deduce $(\delta_{1} + \delta_{2})^{2} \leq
2(\delta_{1}^{2} + \delta_{2}^{2})$ and since $r_i^2 = \rho_i^2 + \delta_i^2$ we
finally obtain 
\[ 
\Delta^{2} \geq \gamma^{2} - (\delta_{1} + \delta_{2})^{2} > 2 (r_1^2+r_2^2) -2
(\delta_1^2 + \delta_{2}^{2}) = 2(\rho_{1}^{2} + \rho_{2}^{2})
\]
and the claim follows.
\end{proof}

The following lemma shows that two pairwise-inflatable balls in
dimension~$d$ can always be ``inflated''\footnote{Hence the name
``pairwise-inflatable''.} to two disjoint equal-radius balls in
dimension~$d+1$.
\begin{lemma}
  \label{inflatability}
  Let $E$ be a $d$-dimensional subspace of $\R^{d+1}$, and let
  $B'_{1}, B'_{2}\subset E$ be pairwise-inflatable $d$-dimensional
  balls in~$E$.  Then there exist two disjoint $(d+1)$-dimensional
  balls $B_{1}, B_{2}$ of \emph{equal radius} in $\R^{d+1}$ such that
  $B'_{1} = B_{1} \cap E$ and $B'_{2} = B_{2}\cap E$.
\end{lemma}
\begin{proof}
  Let $q_{i}$ and $\rho_{i}$ be the center and radius of $B'_{i}$, for
  $i = 1,2$. Consider the line orthogonal to $E$ through $q_i$. Pick a point $p_i$
  on this line at distance $\delta_{i}$ from $q_{i}$, in such a way that $p_{1}$
  and $p_{2}$ are on opposite sides of~$E$. Let also $B_{i}$ be the
  ball with center $p_{i}$ and radius $r_{i} = \sqrt{\delta_{i}^{2} +
  \rho_{i}^{2}}$.  Clearly $B'_{i} = B_{i} \cap E$ and it remains to pick
  $\delta_{i}$ such that $r_{1} = r_{2}$ and $B_{1}$ and $B_{2}$
  are disjoint. 

Let $\Delta$ be the distance between $q_{1}$ and $q_{2}$. Without loss of
generality, we assume $\rho_{1} > \rho_{2}$. Since $\Delta^{2} > 2(\rho_{1}^{2} +
\rho_{2}^{2})$, there exists $\sigma > 0$ such that  
\[ \sigma^{2} < \min\{\Delta^{2} - 2(\rho_{1}^{2} + \rho_{2}^{2}), \, \rho_{1}^{2} - \rho_{2}^{2}\} \] 
and we can define
\[ \delta_{1} = (\rho_{1}^{2} - \rho_{2}^{2} - \sigma^{2})/(2\sigma) \quad \hbox{ and } \quad \delta_{2} = \delta_{1} + \sigma.\]
Now, since $2\sigma \delta_{1} +  \sigma^{2} = \rho_{1}^{2} - \rho_{2}^{2}$ we
have that $\delta_{2}^2 = \delta_{1}^2 + \rho_{1}^{2} - \rho_{2}^{2}$, and it
follows that $B_1$ and $B_2$ have equal radius $r = r_1 = r_2$. Now, the distance $\gamma$ between their
centers satisfies 
\[\gamma^2 = \Delta^2 + (\delta_{1}+\delta_{2})^2 = (\Delta^{2} + 2 \delta_{1}\delta_{2}) + \delta_{1}^{2} + \delta_{2}^{2}.\]
Since 
\[ \Delta^{2} - 2(\rho_{1}^{2} + \rho_{2}^{2}) > \sigma^{2} = (\delta_{2} - \delta_{1})^{2} =  \delta_{1}^{2} + \delta_{2}^{2} -2 \delta_{1}\delta_{2}\] 
it follows that
\[\Delta^{2} + 2 \delta_{1}\delta_{2} > \delta_{1}^{2} + \delta_{2}^{2} + 2(\rho_{1}^{2} + \rho_{2}^{2})\]
and finally
\[\gamma^2 > 2 (\delta_{1}^{2}+ \rho_{1}^{2}) + 2 (\delta_{2}^{2}+ \rho_{2}^{2}) =
4r^2.
\]
 This shows that $B_{1}$ and $B_{2}$ are disjoint.
\end{proof}

Let now $F = (O,x,y,z,w)$ be an orthogonal frame in four-dimensional
space~$\R^4$.  Let $H$ denote the plane $(O,x,y)$, and let $H(z,w)$ be
the translated copy of $H$ going through the point\footnote{By abuse
of notation, we use the letters $z$ and $w$ to label the coordinate
axes and to represent the coordinates of some specific point, the
meaning being clear from the context.} $(0,0,z,w)$. Given two disjoint
convex sets $A$ and $B$ in $\R^4$, we denote by $Q_{AB}^F \subset
\R^2 \times \s^1$ the set of all $(z, w, \alpha)$ such that there is
an oriented line in $H(z,w)$ that intersects~$A$ before $B$ and that
makes an angle~$\alpha$ with the $x$-axis.
\begin{lemma}\label{QAB-4D}
  If $A$ and $B$ are disjoint congruent balls in $\R^4$ then
  $Q_{AB}^F$ is convex for any orthogonal frame $F$ of~$\R^{4}$.
\end{lemma}

We prove this lemma by showing that $Q_{AB}^F$ is the volume under the
graph of a concave function of two variables, which involves showing
that the Hessian of this function is negative definite. We thus follow
the approach of Holmsen et al.~\cite[proof of Lemma 1]{hkl-httlt-03}
but the details (postponed to Appendix~\ref{app}) are more involved.

We proceed to prove the convexity of $\K(\F)$ (but
not yet its strict convexity) for the $3$-dimensional case.
\begin{lemma}\label{only-convex-inflatable}
  Let $\F$ be a sequence of pairwise-inflatable balls in $\R^3$. Then
  $\K(\F)$ is convex.
\end{lemma}
\begin{proof}
  We need to show that for any pair $v_1, v_2 \in \K(\F)$ the great
  circle arc joining them on $\s^2$ lies in $\K(\F)$ (since $\K(\F)$
  is contained in an open hemisphere of $\s^{2}$, there is a unique
  such arc of length less than~$\pi$).  We thus let $\ell_{1},
  \ell_{2}$ be line transversals to $\F$ with directions $v_{1},
  v_{2}$, and pick a plane $H$ parallel to both $\ell_{1}$
  and~$\ell_{2}$.  We embed the $3$-dimensional space as an affine
  3-space of $\R^4$, and equip $\R^4$ with a frame $F = (O,x,y,z,w)$
  such that $\{w=0\}$ is our original $3$-dimensional space, and
  such that $(O,x,y)$ coincides with~$H$.

 For any pair of balls $(B'_1, B'_2)$ from $\F$ with $B'_1 \prec_{\F} B'_2$,
 Lemma~\ref{inflatability} gives us two balls $B_1, B_2 \subset \R^4$ of equal
 radius such that $B'_i =  B_i \cap \{w=0\}$. By Lemma~\ref{QAB-4D},
 $Q_{B_1B_2}^{F}$ is convex and so 
  $Q_{B'_1B'_2}^{F} = Q_{B_1B_2}^{F} \cap \{w = 0\}$ is convex as well.  It
  follows that
  \[
  Q := \bigcap_{A,B \in \F, \, A \prec_\F B} Q_{AB}^{F}
  \]
  is a convex set.

Each point in $Q$ corresponds to a family of parallel and coplanar
lines such that each pair $(A,B)$ in $\F$ is intersected by at least
one of them in the correct order. Helly's theorem (in one dimension)
implies that there is a line transversal to $\F$ in this family and
this transversal is trivially order-respecting. Let $q_{1},\, q_{2}
\in Q$ be the points representing the line transversals $\ell_{1}$ and
$\ell_{2}$. For any direction $v$ on the great circle arc $v_{1}v_{2}$
there is a point $q$ on the segment $q_{1}q_{2}$ whose associated line
transversal has direction $v$.
%
\end{proof}

We now characterize the boundary of~$\K(\F)$.  This will allow us to
show that $\K(\F)$ is not only convex, but even strictly convex.  The
result will then carry over rather effortlessly to arbitrary
dimension.  Recall that $\Ko(\F)$ is the set of directions of
\emph{strict} transversals to~$\F$. The next lemma shows that $\Ko(\F)$ is the interior of $\K(\F)$.

\begin{lemma}\label{single-transversal}
   Let $\F$ be a sequence of disjoint balls in $\R^3$, $v \in
   \s^{2}$ and $D := \bigcap P_{v}(\F)$. Then $v \in \partial
   \K(\F)$ if and only if $D$ is a point and $v \in \INT(\K(\F))$ if
   and only if $D$ has non-empty interior.  
\end{lemma}
\begin{proof}
  Clearly $v \in \K(\F)$ if and only if $D$ is non-empty.  Since
  $P_{v}(\F)$ is a family of discs, $D$ is either empty, a point, or
  has non-empty interior.  If $D$ has non-empty interior, then a small
  perturbation of the direction $v$ cannot cause $D$ to become empty,
  and so $v \in \INT(\K(\F))$.  It remains to show that if $D$ is a
  point, then $v \in \partial\K(\F)$.
  
  We thus assume that $D$ is a point. Let $k \geq 2$ be the number of
  discs that have this point on their boundary, and let $\ell$ be the
  (unique) transversal of $\F$ with direction~$v$.  If $k=2$ then
  $\ell$ lies in a plane separating two balls and there are directions
  $v'$ arbitrarily close to $v$ such that no line transversal with
  direction~$v'$ to these two balls exists (see
  Figure~\ref{perturbation1}). Thus, $v \in \partial\K(\F)$.
  \begin{figure}[!htb]
    \begin{center}
      \includegraphics[width=10cm, keepaspectratio]{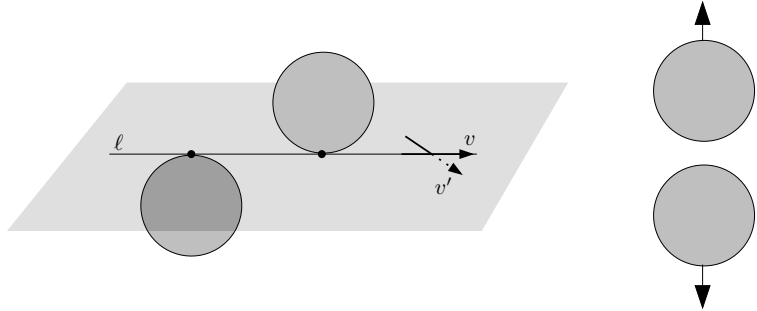}
      \caption{Perturbation removing all transversals when $k=2$: 3D
        view (left) and projections (right). \label{perturbation1}}
    \end{center}
  \end{figure}
  If $k\geq 3$ then by Helly's theorem in the plane there are three
  balls whose projections intersect in a single point. Let $A$ denote
  the middle one with respect to $\prec_\F$ and let $\ell'$ be the
  line through the center of $A$ and its tangency point with~$\ell$
  (see Figure~\ref{perturbation2}). Consider a rotation of $v$ by a
  small angle~$\delta$ around~$\ell'$.
  \begin{figure}[!t]
    \begin{center}
      \includegraphics[width=10cm, keepaspectratio]{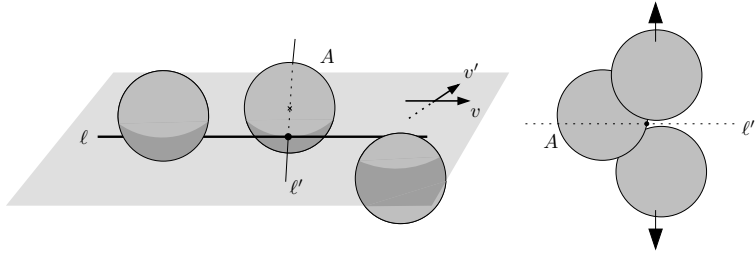}
      \caption{Perturbation removing all transversals when $k=3$: 3D
        view (left) and projections (right).\label{perturbation2}}
    \end{center}
  \end{figure}
  This rotation leaves $P_v(A)$ invariant and moves the centers of the
  two other projections along lines orthogonal to~$P_v(\ell')$, either
  both away from $P_v(\ell')$ or both towards~$P_v(\ell')$, depending
  on the sign of~$\delta$. Any sufficiently small rotation that moves
  the centers away from $P_v(\ell')$ turns $v$ into a direction~$v'$
  such that no transversal to the three balls exists in the
  direction~$v'$. In that case we again have $v \in \partial\K(\F)$.
\end{proof}

\begin{lemma}\label{strict-convexity-3D}
  If $\F$ is a sequence of pairwise inflatable balls in $\R^3$ then
  $\K(\F)$ is strictly convex.
\end{lemma}
\begin{proof}
  We already know that $\K(\F)$ is convex. If $\K(\F)$ is not strictly
  convex then it has to contain on its boundary a great circle arc. By
  the previous lemma, if $v \in \partial \K(\F)$ then $P_v(\F)$ is a
  point. This implies, by Helly's theorem, that the boundary of
  $\K(\F)$ consists of (finitely many) curve arcs that are either (a)
  directions of bitangent lines lying in bitangent planes or (b)
  directions of tritangent lines. The directions of bitangent lines
  lying in bitangent planes to two balls contain a great circle arc
  only if the two balls are tangent, which cannot occur in our
  situation.

  Therefore, if $\K(\F)$ is not strictly convex then it contains in
  its boundary a great circle arc of directions of lines tangent to
  three balls. These directions, being on a great circle arc, are
  parallel to a given plane. In projective geometry, parallels to a
  plane are recast as lines intersecting the ``line at infinity'' of
  that plane. Thus, if $\K(\F)$ is not strictly convex, $\F$ contains
  three balls with infinitely many common tangents that intersect a
  fixed line at infinity. Such configurations were tabulated by
  Megyesi and Sottile~\cite{megyesi05}. Their cases (i), (iii), and (iv)
  cannot arise with disjoint spheres and the fixed line at
  infinity. The remaining possibility (case (ii)) is that the three
  spheres are tangent to a cone whose apex lies on the fixed line. In
  our case, that line is at infinity so this cone is a cylinder and
  the spheres have equal radii and aligned centers; all common
  tangents then have the same direction and cannot form a great circle
  arc.
\end{proof}

We will need the generalization of Lemma~\ref{single-transversal} to
arbitrary dimension.
\begin{lemma}
  \label{single-transversal-d}
  If $\F$ is a sequence of disjoint balls in $\R^d$, then $\Ko(\F) =
  \INT(\K(\F))$.
\end{lemma}
\begin{proof}
  As in the proof of Lemma~\ref{single-transversal} we observe that
  $\Ko(\F) \subset \INT(\K(\F))$, and it remains to prove the other
  inclusion. Let $v \in \INT(\K(\F))$ and pick $v_1, v_2 \in \K(\F)$
  in a neighborhood of $v$ such that $v$ lies in the interior of the
  great circle arc~$v_{1}v_{2}$. Let $\ell_1, \ell_2$ be two line
  transversals to $\F$ with directions $v_1, v_2$, and let $E$ be an
  affine subspace of dimension three containing both lines ($E$ is
  unique if the lines are skew). By Lemma~\ref{lem:pi-closed}, the
  section of $\F$ by $E$ is a sequence $\F'$ of pairwise-inflatable
  balls.  Since $v_{1}$ and $v_{2}$ belong to $\K(\F')$ and $v$ is
  interior to the great circle arc they span,
  Lemma~\ref{strict-convexity-3D} implies that $v \in \INT(\K(\F')) =
  \Ko(\F')$ and, by Lemma~\ref{single-transversal}, there is a
  strict transversal to $\F'$ with direction~$v$. This line is also a
  strict transversal to $\F$ and Lemma~\ref{single-transversal} yields
  that $v \in \Ko(\F)$.
\end{proof}

We can now finally prove the main result of this section.

\begin{proof}[Proof of Proposition~\ref{convex-inflatable}]
  Let $v_1, v_2 \in \K(\F)$ with $v_{1} \neq v_{2}$.  Since $\K(\F)$
  is a closed convex set contained in an open hemisphere of
  $\s^{d-1}$, there is a unique great circle arc of length less
  than~$\pi$ connecting $v_{1}$ and $v_{2}$.  We need to show that all
  interior points of this great circle arc lie in the interior of
  $\K(\F)$.

  Let $\ell_1, \ell_2$ be two line transversals to $\F$ with
  directions $v_{1}$ and~$v_{2}$.  Let $E$ be an affine subspace of
  dimension three containing both transversals.  The space $E$
  intersects every ball in $\F$ and, by Lemma~\ref{lem:pi-closed}, the
  section of $\F$ by $E$ is a sequence $\F'$ of pairwise-inflatable
  balls.
  
  Let $v$ be an interior point of the great circle arc $v_{1}v_{2}$.
  The direction $v$ lies in $E$, and since $\K(\F')$ is strictly
  convex by Lemma~\ref{strict-convexity-3D}, we have $v \in
  \INT(\K(\F')) = \Ko(\F')$.  A strict transversal to $\F'$ is a
  strict transversal to $\F$, and so Lemma~\ref{single-transversal-d}
  implies $v \in \Ko(\F) = \INT(\K(\F))$.
\end{proof}

Proposition~\ref{convex-inflatable} has the following important
corollary:
\begin{cor}\label{empty-interior-anyD}
  Let $\F$ be a sequence of pairwise-inflatable balls in $\R^d$. If
  $\K(\F)$ has empty interior then it is a point.
\end{cor}

\section{Pinning number of pairwise-inflatable balls}

A family $\F$ of objects \emph{pins} a line $\ell$ if $\ell$ is an
isolated transversal to $\F$. The
\emph{pinning number} of a class $\cal C$ of families of objects is
defined as the smallest integer $k$ such that the following holds: if
a family $\F \in \cal C$ pins a line $\ell$ then some subfamily
$\F'\subset \F$ of size at most $k$ already pins $\ell$. A key
ingredient in Hadwiger's original proof of his
theorem~\cite{h-uegt-57} is the fact that the pinning number of
disjoint planar convex sets is $3$. In this section we show a similar
result for pairwise-inflatable balls in $\R^d$. Note that the pinning
number $k$ is simply the Helly number for the property of ``not being
pinned'': if a line transversal to a family $\F$ is not pinned by any
subfamily of size $k$ then it is not pinned by $\F$.

%

\begin{prop} \label{pinning}
The pinning number of pairwise-inflatable balls in $\R^d$ is at most $2d-1$.
\end{prop}

Our proof is based on Lemma~\ref{contractible}, which shows that sets
of compatible transversals are contractible and therefore homology
cells, and Lemma~\ref{weak-hadwiger}, which applies the Topological
Helly Theorem to these sets of lines and obtains a weak version of our
Theorem~\ref{Hadwiger}. We state the next lemma using the notion of
``compatible'' transversal introduced in Section~\ref{sec:prel}:

\begin{lemma}\label{contractible}
  Let $\F$ be a sequence of pairwise-inflatable balls in $\R^d$ and
  $\F'$ be a subsequence of $\F$. Then the set $L$ of line
  transversals to $\F'$ compatible with $\F$ is a contractible subset
  of $\R^{2d-2}$.
\end{lemma}

Note the restriction on the direction of lines in~$L$: there may be
strict order-respecting line transversals to $\F'$ that are not
compatible with $\F$.
\begin{proof}
  Given a line $\ell \in L$, let $v_\ell$ be its direction. A
  transversal $\ell$ to~$\F'$ is \emph{barycentric} if it goes
  through the center of mass of the intersection of
  $P_{v_{\ell}}(\F')$. For any direction $v$ in $\K(\F')$ there is
  a unique barycentric transversal to $\F'$, which we
  denote~$b_{\F'}(v)$.

  Let $L^*$ denote the set of barycentric transversals to
$\F'$ with directions in $\DS$. The projection of a ball changes
  continuously with the direction of projection, so $b_{\F'}$ is
  continuous. Since the direction of a line changes continuously with
  the line, $b_{\F'}^{-1}$ is also continuous. Thus, $b_{\F'}$
  defines a homeomorphism between $L^*$ and~$\K(\F')\cap \DS$.

  By Lemma~\ref{convex-inflatable}, $\K(\F')$ is convex and so is
  $\DS$. Thus, $\K(\F')\cap \DS$ is convex and hence
  contractible. It follows that $L^*$ is also contractible. The map
  \[\left\{
  \begin{array}{l}
    L \times [0,1] \rightarrow L\\
    (\ell,t) \mapsto \ell+t(b_{\F'}(v_{\ell})-\ell)
  \end{array}
  \right.\]
  is continuous and shows that $L^*$ is a deformation retract of
  $L$. Since $L^*$ is contractible, so is $L$.
\end{proof}

We can now apply the Topological Helly Theorem to obtain a
``weak'' Hadwi\-ger-type result.

\begin{lemma}\label{weak-hadwiger}
  Let $\F$ be a sequence of at least $2d-1$ pairwise-inflatable balls
  in~$\R^d$. If every subfamily $\F'\subset \F$ of $2d-1$ balls admits
  a strict line transversal with a direction in $\DS$, then $\F$
  admits a strict line transversal.
\end{lemma}
\begin{proof}
  We apply Theorem~\ref{thm:topological-helly} on $\Lines(\DS)$. With
  the parametrization discussed above, $\Lines(\DS) \subset
  \R^{2d-2}$.  For $S \in \F$ let $X_{S}$ be the subset of
  $\Lines(\DS)$ of lines intersecting the interior of ball~$S$.
  Clearly, $X_{S}$ is an open set in~$\R^{2d-2}$.  Consider now the
  intersection $Y := X_{S_{1}}\cap \dots \cap X_{S_{r}}$ of $r$ such
  sets.  The set $Y$ consists of exactly those lines in $\Lines(\DS)$
  that are strict transversals of $S_{1},\dots, S_{r}$. The assumption
  of the lemma implies that $Y \neq \emptyset$ for $r \leq 5$. By
  Lemma~\ref{contractible}, $Y$ is contractible and hence a homology
  cell. Theorem~\ref{thm:topological-helly} now implies that
  $\bigcap_{S\in\F} X_{S} \neq \emptyset$, and so there is an
  order-respecting strict line transversal for~$\F$.
\end{proof}

In principle, Lemma~\ref{weak-hadwiger} is the Hadwiger-type result we
are looking for.  Its drawback is that it requires a subfamily of
balls to have not only an order-respecting transversal, but one that,
in a sense, respects the order on the \emph{entire} family of balls.
This is nonetheless enough to prove the desired result on the pinning number
of pairwise-inflatable balls:

\begin{proof}[Proof of Proposition~\ref{pinning}]
  Let $\F$ be a family of at least $2d$ pairwise-inflatable balls
  in~$\R^d$ admitting an isolated line transversal~$\ell$.  Let
  $\prec$ be the order on $\F$ induced by~$\ell$.
  Lemma~\ref{contractible} implies that the set of line transversals
  to $\F$ respecting $\prec$ is connected, and so $\ell$ is the only
  order-respecting line transversal to~$\F$.

  Since $\ell$ is not a strict transversal, $\F$ has no strict
  order-respecting transversal. By Lemma~\ref{weak-hadwiger}, there is
  a subfamily $\F'\subset \F$ of $2d-1$ balls that has no strict
  order-respecting transversal with direction in~$\DS$, that is
  $\Ko(\F')\cap\DS=\emptyset$. However, $\K(\F')\cap\DS \neq
  \emptyset$ since it contains the direction of $\ell$. Since
  $\K(\F')$ is convex, by Lemma~\ref{convex-inflatable}, and $\DS$ is
  open, it follows that $\Ko(\F') = \emptyset$ and $\F'$ has no strict
  order-respecting transversal at all. Now, $\K(\F')$ is non-empty but
  has empty interior, so, by Corollary~\ref{empty-interior-anyD},
  $\K(\F')$ is a single direction $v$. Since $\K(\F') = \{v\}$, the
  balls $P_v(\F')$ intersect in a unique point and $\ell$ is the only
  order-respecting line transversal of $\F'$, and is thus isolated.
\end{proof}

\section{Hadwiger and Helly-type theorems}
\label{hadwiger-and-helly}

We can now prove the main results of this paper.

\paragraph{A Hadwiger-type theorem.}
Propositions~\ref{empty-interior-anyD} and~\ref{pinning} are all we
need to reproduce Hadwiger's original proof of the $2$-dimensional
case.

\begin{proof}[Proof of Theorem~\ref{Hadwiger}]
  We simultaneously shrink all the balls and continue shrinking as long as every
  subset of size $2d$ has a transversal. If all the centers are aligned then the
  theorem trivially holds. Otherwise, at some point in the shrinking process a
  subfamily $\F'$ of size $2d$ stops having a transversal. The cone $\K(\F')$
  changes continuously during the shrinking and must have empty interior before
  disappearing. Thus, by Corollary~\ref{empty-interior-anyD}, at that moment the
  sequence $\F'$ has a unique transversal $\ell$.

  Now, by Proposition~\ref{pinning}, there is then a subfamily $\F''
  \subset \F'$ of at most $2d-1$ balls such that $\ell$ is the unique
  transversal of~$\F''$.  For any ball $X \in \F \setminus \F''$, the
  set $\F'' \cup \{ X \}$ has a line transversal~$\ell_X$.  Since the
  only line transversal of $\F''$ is~$\ell$, we must have $\ell_X =
  \ell$, and $\ell$ intersects~$X$. It follows that $\ell$ is an
  unordered line transversal for~$\F$.

  Similarly, if any subfamily of size $2d+1$ admits a line transversal
  there exists a subfamily $\F'$ of $2d-1$ balls having a unique line
  transversal~$\ell$. For any $X,Y \in \F$ with $X \prec Y$, the
  subfamily $\F' \cup \{X,Y\}$ admits a line transversal that must
  be~$\ell$, and $\ell$ intersects $X$ before~$Y$. It follows that
  $\ell$ is an (order-respecting) line transversal of $\F$.
\end{proof}

\paragraph{Removing the ordering assumption.} We now generalize
Theorem~\ref{Hadwiger} by removing the restriction on the ordering.
However, we restrict ourselves to the case of disjoint unit balls in
$\R^d$ as we build on the following result by Cheong et
al.~\cite{cgn-gpdus-05}.
\begin{theo}[\cite{cgn-gpdus-05}]
  \label{thm:gp}
  Let $\F$ be a family of at least nine disjoint unit balls in $\R^d$.
  Then $\F$ admits at most two distinct geometric permutations, which
  differ only in the swapping of two adjacent balls.
\end{theo}

\begin{proof}[Proof of Theorem~\ref{Helly-type}]
  We first shrink the balls simultaneously until some subfamily
  $\F_{4d-1}$ of $4d-1$ balls is about to lose its last unordered transversal.

  If $\F_{4d-1}$ admits more than one (unordered) line transversal
  (all of which vanish if the balls are shrunk any further), each
  transversal must realize a different geometric
  permutation. Theorem~\ref{thm:gp} then implies that $\F_{4d-1}$ has
  exactly two line transversals, $\ell_1$ and $\ell_2$, with two
  distinct geometric permutations. By Proposition~\ref{pinning}, for each
  $\ell_i$ there are $2d-1$ balls in $\F_{4d-1}$ for which $\ell_i$ is
  the only line transversal respecting the ordering induced by
  $\ell_{i}$.  There is thus a subfamily $\F'$ of $\F_{4d-1}$ of
  exactly $4d-2$ balls (we can complete $\F'$ using balls from
  $\F_{4d-1}$ if needed) for which $\ell_1$ and $\ell_2$ are the only
  line transversals respecting their respective orders.  By
  Theorem~\ref{thm:gp}, $\F'$ admits at most two geometric
  permutations, and so $\ell_1$ and $\ell_2$ are its only line
  transversals. Since any subfamily of $4d-1$ balls has a line
  transversal, any ball of $\F \setminus \F'$ must intersect $\ell_1$
  or $\ell_2$. If all the balls intersect both lines then the theorem
  is proved. Otherwise, there exists a ball $A$ that intersects, say,
  $\ell_1$ but not $\ell_2$. Then $\F'\cup\{A\}$ is a family of $4d-1$
  balls with a \emph{unique} transversal.  We are left with a set
  $\F_{4d-1}$ of $4d-1$ balls that has a unique transversal~$\ell$.

  Let $\prec_{\ell}$ be the order on $\F_{4d-1}$ induced by~$\ell$. By
  Proposition~\ref{pinning}, there is a subfamily $\F_{2d-1} \subset
  \F_{4d-1}$ such that $\ell$ is the unique transversal of $\F_{2d-1}$
  respecting~$\prec_{\ell}$. For each $Z \in \F_{4d-1} \setminus
  \F_{2d-1}$, let $\F_{Z}$ denote the set $\F_{4d-1} \setminus \{Z\}$.
  If one of the subsets $\F_Z$ has no other transversal than $\ell$
  then every other ball of $\F$ intersects $\ell$ and the proof is
  complete.

  We now assume that every $\F_Z$ has some transversal $\ell_Z$
  distinct from $\ell$ and obtain a contradiction. Since $\F_Z$
  contains $\F_{2d-1}$, $\ell_{Z}$ realizes a geometric permutation
  different from that of $\ell$. By Theorem~\ref{thm:gp}, the order
  induced by $\ell_{Z}$ on $\F_{4d-1}$ differs from~$\prec_{\ell}$ by
  the swapping of two adjacent balls~$X,Y$. Since $\ell_{Z}$ realizes
  a geometric permutation of $\F_{2d-1}$ different from $\ell$, we
  must have $X,Y \in \F_{2d-1}$.  Let $Z_{1}, Z_{2} \in \F_{4d-1}
  \setminus \F_{2d-1}$, and consider the set $\F_{4d-1} \setminus
  \{Z_{1},Z_{2}\}$. It admits the transversals $\ell$, $\ell_{Z_{1}}$,
  and $\ell_{Z_{2}}$ but, by Theorem~\ref{thm:gp}, at most two
  geometric permutations. Since $\ell$ is the unique transversal
  respecting~$\prec_{\ell}$, $\ell_{Z_1}$ and $\ell_{Z_2}$ must
  realize the same geometric permutation on $\F_{4d-1} \setminus
  \{Z_{1}, Z_{2}\}$. Thus the balls $X,Y \in \F$ do not depend on the
  choice of~$Z$.  Let $\prec$ be the order on $\F_{4d-1}$ obtained
  from $\prec_{\ell}$ by swapping $X$ and~$Y$.  For any $Z \in
  \F_{4d-1} \setminus \F_{2d-1}$ the subfamily $\F_{Z}$ admits a line
  transversal respecting~$\prec$. On the other hand, $\F_{4d-1}$ does
  not admit such a transversal as $\ell$ is its only transversal. By
  (the second half of) Theorem~\ref{Hadwiger}, there is a subset
  $\F_{2d+1}\subset \F_{4d-1}$ of at most $2d+1$ balls that does not
  admit a transversal respecting~$\prec$.  We must have $X,Y \in
  \F_{2d+1}$, as without both $X$ and $Y$, $\prec_{\ell}$ and $\prec$
  are equivalent.  This implies that $|\F_{2d-1} \cup \F_{2d+1}| \leq
  4d-2$.  There is therefore a $Z \in \F_{4d-1} \setminus \F_{2d-1}$
  such that $\F_{2d-1} \cup \F_{2d+1} \subseteq \F_{Z}$. However,
  $\ell_{Z}$ cannot be a line transversal to $\F_{2d+1}$, a
  contradiction.
\end{proof}

\section{Conclusion and open problems}

We conclude this paper with a few comments on our results followed by
open problems they suggest.

\begin{itemize}
\item Weaker versions of Theorems~\ref{Hadwiger} and~\ref{Helly-type}
  (with constants quadratic in $d$) can be obtained more easily, using
  only Lemma~\ref{convex-inflatable} and 
  the reasoning of Holmsen et al.~\cite{hkl-httlt-03}.

\item In the plane, if three disjoint convex sets $\{C_1, \ldots,
  C_3\}$ pin a line $\ell$ then they are all tangent to $\ell$ and
  alternate: the first and the third are on the same side of $\ell$,
  the second is on the other side. Thus, if $\ell$ does not intersect
  a fourth convex set $C_4$ some triple $\{C_x, C_y, C_4\}$ has no
  line transversal at all. This explains why, in Hadwiger's original
  proof the ``Hadwiger number'' is the same as the pinning number. A
  way to reduce the bound in Theorem~\ref{Hadwiger} to~$2d-1$
  could be to prove a similar statement: given a sequence of pairwise
  inflatable balls $\F$ that pins a line $\ell$ and a ball $C$ not
  intersecting $\ell$, there is a subsequence $\F' \subset \F$ of size
  $|\F|-1$ such that $\F' \cup \{C\}$ has no transversal respecting
  the ordering on $\F'$. We have no idea whether such a statement
  actually holds.

\item To apply the Topological Helly Theorem, we did not actually need
  that $\K(\F)$ is convex, only that it is contractible.  This may be
  important for further generalization.

\item For general convex sets, even smooth ones, the pinning number is
  at least $6$ as the following example using six unit-radius
  cylinders in $\R^3$, due to G\"unter Rote, shows: the first three
  cylinders are parallel to the $x$-axis and their axes go through the
  points $(0,1,0)$, $(0,-1,1)$ and $(0,1,2)$ respectively. The last
  three cylinders are parallel to the $y$-axis and their axes go
  through the points $(1,0,10)$, $(-1,0,11)$ and $(1,0,12)$
  respectively. The six cylinders have only one transversal---the
  $z$-axis---but any five have an infinite number of transversals.

\item Lemmas~\ref{lem:pi-closed} and~\ref{inflatability} imply that
  two disjoint balls $A,B \subset \R^{d}$ are pairwise-inflatable if
  and only if they can be expressed as sections of two disjoint
  congruent balls in some higher-dimensional space.  Generalizing
  this, let us call a set $\F$ of balls in $\R^{d}$ \emph{inflatable}
  if $\F$ can be expressed as the intersection of a higher-dimensional
  set of disjoint congruent balls with a $d$-dimensional affine
  subspace.  Batog recently showed that it is NP-hard to decide
  whether a given collection of balls is inflatable~\cite{batog05}.
\end{itemize}

\bigskip

\begin{pb}
  What is the maximum number of geometric permutations of
  pairwise-inflatable balls in $\R^d$?
\end{pb}
To generalize Theorem~\ref{Helly-type} to
pairwise-inflatable balls, one would need to extend
Theorem~\ref{thm:gp} to those families. It is known that the number of
geometric permutations of $n$ disjoint balls in $\R^d$ is at most $3$
if the balls have equal radii and $\Theta(n^{d-1})$ if the ratio
\[
\frac{\hbox{largest radius}}{\hbox{smallest radius}}
\]
is not bounded independently of $n$~\cite{zs-gpbbs-03}. 

\medskip
\begin{pb}\label{pb-cone}
  For which classes of objects is the cone of directions $\K(A_1,
  \ldots, A_n)$ convex, or at least contractible?
\end{pb}
Our proof of convexity for the cone of directions of balls collapses
for balls that are not pairwise-inflatable. In fact, the set $Q_{AB}^{F}$ is
not necessarily convex if $B$ is much smaller than $A$ but very close
to it.

\begin{pb}\label{pb-transversals}
  For which classes of objects is the set of order-respecting line
  transversals always connected?
\end{pb}
Our proof of Theorem~\ref{Hadwiger} follows
from (i) a bounded pinning number and (ii) the fact that as the set of
order-respecting line transversals to a sequence disappears it first
reduces to a single line. For strictly convex objects, property (ii)
follows from the connectivity of the set of order-respecting
transversals. Surprisingly, it is an open question whether this set is
connected for even $4$ disjoint balls in $\R^3$, whereas it is known
to be connected for any triple of disjoint convex objects~\cite[Lemma
74]{g-svgtcd-04}. We conjecture that general convex sets in $\R^d$
have a bounded pinning number. Thus, understanding how general this
connectivity property is would provide insight in how general the
example of Holmsen and Matousek~\cite{hm-nhtst-04}, convex sets whose
translates do not admit a Hadwiger theorem, actually is. Of course, a
positive answer to Problem~\ref{pb-cone} for a particular family of
convex sets implies a positive answer to Problem~\ref{pb-transversals}
for that family as well.

\begin{pb}
  Given a collection of disjoint unit balls, assume that any subset of
  size $2d-1$ admits a line transversal. Does any subset of size
  $2d-1$ admit a \emph{compatible} line transversal?
\end{pb}
In other words, can our ``weak Hadwiger theorem'' 
(Lemma~\ref{weak-hadwiger}) be strengthened into a Hadwiger theorem with
a better constant than Theorem~\ref{Hadwiger}?

\begin{pb}
  Is the pinning number of disjoint unit balls in $\R^d$ equal to $2d-1$?
\end{pb}
Surprisingly, the only known lower bound on the Helly number is the
construction done by Hadwiger fifty years ago. Note that the bound in
our Hadwiger theorem has to be higher than the pinning number of the
corresponding family and one can therefore look for a lower bound on
the pinning number. Intuitively, considerations on the dimension
suggest that the pinning number in dimension $d$ cannot be less
than~$2d-1$, the dimension of the underlying line space being~$2d-2$.

\section*{Acknowledgments}
We thank Gregory Ginot for helpful discussions and suggesting the
proof of Lemma~\ref{contractible}, G\"unter Rote for the lower bound
construction with cylinders mentioned in the conclusion, and Guillaume
Batog for helpful discussions on inflatability.

\bibliographystyle{abbrv}
\bibliography{transversal}

\appendix

\section{Proof of Lemma~\ref{QAB-4D}\label{app}}

\begin{proof}
  Let $F$ be the frame $(O, x, y, z, w)$.  We first observe that a
  translation of~$F$ along the $x$- or $y$-axis leaves $Q_{AB}^{F}$
  unchanged, while a translation of~$F$ along the $z$- or $w$-axis
  causes an equivalent translation of $Q_{AB}^{F}$.  Rotating the $x$-
  and $y$-axes while leaving the $z$- and $w$-axes fixed causes a
  translation of $Q_{AB}^{F}$ along the $\alpha$-axis.  Finally,
  scaling $F$ causes $Q_{AB}^{F}$ to be stretched along the $z$- and
  $w$-axes.  Since convexity is invariant under affine
  transformations, we can therefore assume that $A$ and $B$ are
  \emph{unit-radius} balls with centers at $(0,0,0,-b)$ and
  $(e,0,0,b)$, where~$b > 0,\, e > 0$.  The disjointness of $A$ and
  $B$ implies that $e^2+4b^2-4 > 0$. Let $D$ denote the lune-shaped
  region in the $(z,w)$ plane that corresponds to the intersection of
  the two unit discs with centers $(0,-b)$ and $(0,b)$. If $(z,w)
  \notin D$ then $H(z,w)$ does not intersect both $A$ and $B$.  If $b
  > 1$ then $D$ is empty. If $b = 1$ then $D$ is reduced to $z = w =
  0$, $H(0,0)$ intersects both $A$ and $B$ in a point, and so
  $Q_{AB}^F$ is a point.  In the following we can therefore assume~$b
  < 1$.

  Let
  \[ 
  R(z,w) = \sqrt{1 - z^2 - w^2},
  \] 
  and let $R_+ = R(z,w+b)$ and $R_- = R(z,w-b)$. If $(z,w) \in D$ then
  $H(z,w) \cap A$ is the disc with center $(0,0)$ and radius~$R_+$,
  while $H(z,w) \cap B$ is the disc with center $(0,e)$ and
  radius~$R_-$.  Now, let
  \[ 
     f(z,w) = \frac{R_+ + R_-}{e}.
  \]
  Since $A$ and $B$ are disjoint, the discs $H(z,w) \cap A$ and
  $H(z,w) \cap B$ are disjoint, implying that $R_+ + R_- < e$, and so
  $0 \leq f(z,w) < 1$. Consider
  \[
     G(z,w) = \arcsin(f(z,w)).
  \] 
  Since $(z,w,\alpha) \in Q_{AB}^{F}$ if and only if $(z,w) \in D$ and
  $-G(z,w) \leq \alpha \leq G(z,w)$, it suffices to show that $G$ is a
  concave function.  A sufficient condition for this is that its
  Hessian $\h(G)$ be negative definite, which we endeavor to prove
  now. By symmetry with respect to the $z$- and $w$-axes, we need to
  prove negative definiteness only for~$z,\,w \geq 0$.

  In what follows, subscripts are used to denote partial
  derivatives. Also, reference to $z,w$ as arguments of functions is
  dropped when no confusion can arise.

  The Hessian of $G$ is
  \[ 
     \h(G) = \begin{pmatrix} G_{zz} & G_{zw}\\ G_{zw} & G_{ww}\end{pmatrix}
     = \frac{(1-f^2)\h(f)+ f (\nabla f) (\nabla f)^T}{(1-f^2)^{3/2}}
  \]
  where $\h(f)$ is the Hessian of $f$ and $\nabla f = (f_z,f_w)^T$ is
  its gradient. The Hessian of $G$ is negative definite if and only if
  \[ 
     \text{(i)} \quad G_{zz} < 0 \quad \hbox{and} \quad
     \text{(ii)} \quad \det{\h(G)} = G_{zz} G_{ww} - G_{zw}^2 > 0.
  \]
  We prove these two inequalities in turn. For this, we need the following
  derivatives:
  \[
     R_z = \frac{-z}{R}, \quad R_w = \frac{-w}{R}, \quad R_{zz} = 
     \frac{w^2-1}{R^3}, 
     \quad R_{zw} = \frac{-zw}{R^3}, \quad R_{ww} = \frac{z^2-1}{R^3}, 
     \quad R_{zzz} = \frac{3(w^2-1)z}{R^5}
  \]

  \bigskip
  \noindent (i). The first inequality is simple to check. We have
  \[ 
     G_{zz} = \frac{(1-f^2)f_{zz}+f \, f_z^2}{(1-f^2)^{3/2}}.
  \]
  Since the denominator is strictly positive for all $z$ and $w$, the
  sign of $G_{zz}$ is determined by its numerator which we denote
  by $g(z,w)$. The derivative of $g$ with respect to $z$ is:
  \[ 
     g_z = (1-f^2)f_{zzz}+f_z^3.
  \] 
  For $z > 0$, we have $R_z < 0$ and $R_{zzz} < 0$, so $f_z < 0$ and
  $f_{zzz} < 0$ implying that $g_z < 0$. It follows that the function $z
  \mapsto g(z,w)$ is decreasing for $z > 0$. Since $g(0,w) < 0$ it
  follows that $g(z,w) <0$ for $z,w \geq 0$, so $G_{zz} < 0$.

  \bigskip
  \noindent
  (ii). The second inequality is considerably more challenging. Let us
  introduce the following notations:
  \begin{gather*}
    \gamma_+ = R_+^2, \quad \gamma_- = R_-^2, \quad \gamma = 1-z^2-w^2+b^2,\\
    P = \gamma_+ \gamma_-, \quad S = \gamma_+ +\gamma_-.
  \end{gather*}
  $\gamma_+, \gamma_-$ and $\gamma$ satisfy the following constraints:
  \[
     0 < \gamma_+ \leq 1-b^2 < 1, \quad
     0 < \gamma_- \leq 4b(1-b) < 1 \quad \hbox{and} \quad
     0 < 2b^2 \leq \gamma < 1+b^2 < 2.
  \]
  Expanding $\det{\h(G)}$ gives $\det{\h(G)} = (1-f^2) \Delta$, where
  \begin{gather*}
     \Delta = (1-f^2) \Delta_1 + f \Delta_2,\\
     \Delta_1 = \det{\h(f)} = f_{zz} f_{ww} - f_{zw}^2, \quad
     \Delta_2 = f_w^2 f_{zz} + f_z^2 f_{ww} - 2f_z f_w f_{zw}.
  \end{gather*}

  We first find that
  \[
     \Delta_1 = \frac{1}{e^2 P^2} (\mu_1 + \mu_2 \sqrt{P}),
  \]
  where
  \[
     \mu_1 = S^2-2P = \gamma_-^2 + \gamma_+^2 > 0 \quad \hbox{and} \quad
     \mu_2 = P + \gamma(2-\gamma) > 0.
  \]
  Also,
  \[
     \Delta_2 = \frac{1}{e^3 P^\frac{3}{2}} (\lambda_- \sqrt{\gamma_-} 
     + \lambda_+ \sqrt{\gamma_+}),
  \]
  where
  \[
     \lambda_- = \gamma (\gamma-2) + 2 \gamma_+ (\gamma-1) 
     + P \quad \hbox{and} \quad 
     \lambda_+ = \gamma (\gamma-2) + 2 \gamma_- (\gamma-1) + P.
  \]
  Note that since $\lambda_-(z,0) = \lambda_+(z,0) = 4z^2(z^2-1) \leq 0$,
  we can't conclude yet and have to go further along. 

  Putting everything together, we get
  \[
     \Delta = \frac{\chi}{e^4 P^2},
  \]
  where 
  \begin{gather*}
    \chi = \chi_1 + \chi_2 \sqrt{P},\\
    \chi_1 = \mu_1 (e^2-S) + P (\lambda_+ + \lambda_- -2\mu_2), \quad
    \chi_2 = \mu_2(e^2-S) -2\mu_1 + \lambda_- \gamma_- + \lambda_+ \gamma_+.
  \end{gather*}
  We want to prove that $\chi > 0$, implying $\Delta > 0$. Let $\delta =
  e^2+4b^2-4$. Noting that $S+ 4-2\gamma = \gamma_+ + \gamma_- +4- 2\gamma
  = 4-4b^2$, we get that $e^2-S = \delta+4-2\gamma$. So we have:
  \[
    \chi_1 = \mu_1 \delta + \chi_1^*, \quad \chi_2 = \mu_2 \delta + \chi_2^*,
  \]
  where
  \begin{gather*}
    \chi_1^* = 2 \mu_1(2-\gamma) + P (\lambda_+ + \lambda_- -2\mu_2),\\
    \chi_2^* = -2\mu_1 + 2\mu_2(2-\gamma) + \lambda_- \gamma_- 
    + \lambda_+ \gamma_+.
  \end{gather*}
  Let $\chi^* = \chi_1^* + \chi_2^* \sqrt{P}$. Then
  \[
     \chi = (\mu_1 + \mu_2 \sqrt{P}) \delta + \chi^* > \chi^*,
  \]
  since $\mu_1 > 0, \mu_2 > 0, \delta > 0$. 

  Let us prove that $\chi^* \geq 0$. Let
  \[
     \theta_1 = 2S^2-4P-SP-2P\gamma, \quad \theta_2 = 2(2-\gamma)-S.
  \]
  We can rewrite $\chi_1^*$ and $\chi_2^*$ in terms of $\theta_1$ and
  $\theta_2$:
  \[
     \chi_1^* = (2-\gamma) \theta_1 - P\gamma \theta_2, \quad
     \chi_2^* = -\theta_1 + \gamma (2-\gamma) \theta_2.
  \]
  Now observe that $\chi^*$ factors:
  \[
     \chi^* = \chi_1^* + \chi_2^* \sqrt{P} =  
     (2-\gamma-\sqrt{P})(\theta_1 + \theta_2 \gamma \sqrt{P}).
  \]
  Noting that $\theta_2 = 4(w^2+z^2) \geq 0$ and
  \[
     \theta_1 = 2S^2-8P + P(2(2-\gamma)-S) = 2 (\gamma_+-\gamma_-)^2
       + P \theta_2 \geq 0,
  \]
  we see that the second factor of $\chi^*$ is positive. It remains to
  observe that $2-\gamma + \sqrt{P} > 0$ and that
  \[
     (2-\gamma)^2 - P = 4 (z^2(1-b^2) + w^2) \geq 0,
  \]
  to conclude that $2-\gamma-\sqrt{P} \geq 0$ and $\chi^* \geq 0$. Overall,
  $\chi > 0, \Delta > 0$ and $\det{\h(G)} > 0$, which concludes the proof.
\end{proof}

\end{document}